\documentclass[lettersize,journal]{IEEEtran}
\usepackage{amsmath,amsfonts}
\usepackage{algorithm}
\usepackage{array}
\usepackage[caption=false,font=normalsize,labelfont=sf,textfont=sf]{subfig}
\usepackage{textcomp}
\usepackage{stfloats}
\usepackage{url}
\usepackage{verbatim}
\usepackage{graphicx}

\usepackage{multirow}%
\usepackage{amssymb}%
\usepackage{amsthm}%
\usepackage{mathrsfs}%
\usepackage{xcolor}%
\usepackage{textcomp}%
\usepackage{manyfoot}%
\usepackage{booktabs}%
\usepackage{algorithmicx}%
\usepackage{algpseudocode}%
\usepackage{listings}%

\begin{document}

\title{Towards General Auditory Intelligence: Large Multimodal Models for Machine Listening and Speaking}

% \title[Article Title]{Towards General Auditory Intelligence: Large Multimodal Models for Machine Listening and Speaking}

%\author{Siyin Wang$^{1,*}$, Zengrui Jin$^{1,*}$, Changli Tang$^1$, Wenyi Yu$^1$, Yixuan Li$^1$,  Jimin Zhuang$^1$, Yudong Yang$^1$, Guangzhi Sun$^5$, Ji Wu$^1$, Bowen Zhou$^1$, Chao Zhang$^{1,\S}$,  Yuchen Hu$^2$, Xiaohai Tian$^2$, Xianzhao Chen$^2$, Jun Zhang$^2$, Lu Lu$^2$, Yuxuan Wang$^2$, Yonghui Wu$^2$, Chen Chen$^3$, Zhehuai Chen$^3$, Mingqiu Wang$^4$, Qiujia Li$^4$, Michael Han$^4$, Yifan Ding$^4$, Junwen Bai$^4$, Tom Ouyang$^4$, Shuo-yiin Chang$^4$, Bo Li$^4$, Tara Sainath$^4$

%\author{S. Wang$^{1,*}$, Z. Jin$^{1,*}$, C. Tang$^1$, W. Yu$^1$, Y. Li$^1$, J. Zhuang$^1$, Y. Yang$^1$, G. Sun$^5$, J. Wu$^1$, B. Zhou$^1$, C. Zhang$^{1,\S}$\\ 
%Y. Hu$^2$, X. Tian$^2$, X. Chen$^2$, J. Zhang$^2$, L. Lu$^2$, Y. Wang$^2$, Y. Wu$^2$, C. Chen$^3$, Z. Chen$^3$\\
%M. Wang$^4$, Q. Li$^4$, M. Han$^4$, Y. Ding$^4$, J. Bai$^4$, T. Ouyang$^4$, S.-y. Chang$^4$, B. Li$^4$, T. Sainath$^4$

\author{Siyin Wang$^{1,*}$, Zengrui Jin$^{1,*}$, Changli Tang$^1$, Qiujia Li$^2$, Bo Li$^2$, Chen Chen$^3$, Yuchen Hu, Wenyi Yu$^1$, 
Yixuan Li$^1$, Jimin Zhuang$^1$, Yudong Yang$^1$, Mingqiu Wang$^2$, Michael Han$^2$, Yifan Ding$^2$, Junwen Bai$^2$, 
Tom Ouyang$^2$, Shuo-yiin Chang$^2$, Xianzhao Chen, Xiaohai Tian, Jun Zhang, Lu Lu, Guangzhi Sun$^4$,
Zhehuai Chen$^3$, Ji Wu$^1$, Bowen Zhou$^1$, Yuxuan Wang, Tara Sainath$^2$, Yonghui Wu, Chao Zhang$^{1,\S}$\\
%        % <-this % stops a space
%% \thanks{This paper was produced by the IEEE Publication Technology Group. They are in Piscataway, NJ.}% <-this % stops a space
%%\\
%

\ \\

$^1${Tsinghua University}, $^2${Google DeepMind},
$^3${NVIDIA}, $^4${University of Cambridge} \\
$*$ Equal contribution, $\S$ Corresponding author: \texttt{cz277@tsinghua.edu.cn}
}

\maketitle

\begin{abstract}
\ In the era of large language models (LLMs) and artificial general intelligence (AGI), computer audition must evolve beyond traditional paradigms to fully leverage the capabilities of foundation models, towards more comprehensive understanding, more natural generation and more human-like interaction. Audio, as a modality rich in semantic, emotional, and contextual cues, plays a vital role in achieving naturalistic and embodied machine intelligence. This survey provides a comprehensive review of recent progress in integrating audio into LLMs, with a focus on four key areas: audio comprehension, audio generation, speech-based interaction, and audio-visual understanding. We analyze how LLMs are reshaping audio perception and reasoning, enabling systems to understand sound at a deeper semantic level, generate expressive audio outputs, and engage in human-like spoken interaction. Furthermore, we explore how the fusion of audio and visual modalities enhances situational awareness and cross-modal reasoning, pushing the boundaries of multimodal intelligence. This survey not only synthesizes existing research but also identifies critical challenges and future directions for building audio-native AGI systems capable of perceiving, understanding, and interacting through sound as naturally as humans do. 
\end{abstract}

\section{Introduction}
In the rapidly evolving field of artificial intelligence, large language models (LLMs) have demonstrated exceptional proficiency in processing and generating text sequences for both natural and formal languages. Models scaled to billions of parameters, such as ChatGPT, Gemini, Deepseek-R1 and LLaMA \cite{touvron2023llama}, have established new benchmarks in general-purpose language understanding and few-shot learning capabilities. Building upon this foundation, current artificial intelligence (AI) research is increasingly extending LLMs to incorporate additional modalities, including but not limited to audio, images, and videos, resulting in Multimodal LLMs. Many recent studies have focused on integrating audio into these models, utilizing the advanced comprehension capabilities of LLMs to address various audio understanding \cite{audiopalm,zhang-etal-2023-speechgpt,salmonn} and generation tasks \cite{cosyvoice,yang2023uniaudio,wang2024speechx}, to develop generalized audio processing capabilities.
A critical milestone in this evolution is the emergence of omni-modal models such as GPT-4o \cite{gpt4o}, Gemini and gpt-realtime \cite{gptrealtime}, which integrate audio interactions encompassing diverse emotional expressions and tonal variations. This advancement marks significant progress towards developing AI systems that combine human-like audio-visual perception and language cognition abilities. The capability of AI systems to perceive and interpret a broad range of auditory signals is crucial for numerous real-world applications, particularly because it uniquely provides critical context, emotional nuances, and semantic depth that is often inaccessible through visual or text data alone, making it a fundamental modality for creating truly versatile AI assistants capable of natural human interactions. % This drives toward unified multimodal AI agents promises significant advancements in the human-like qualities of future AI systems.

\begin{figure}[!tp]
\centering
\includegraphics[width=\linewidth]{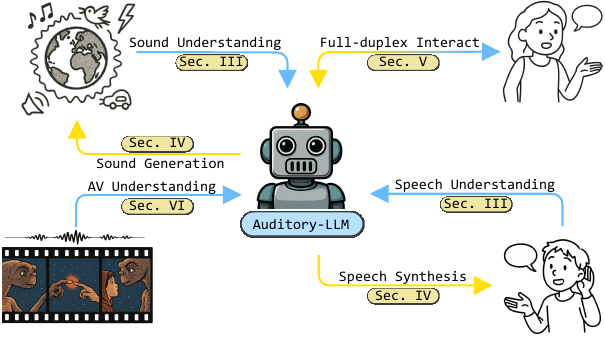}
\caption{Overview of how an Auditory Large Language Model (Auditory-LLM) interfaces with the world and humans through the audio modality. It processes diverse environmental and speech sounds via audio understanding, generates naturalistic outputs through audio and speech synthesis, and enables real-time full-duplex interaction (simultaneously listening and responding), enabling seamless auditory communication.}
\end{figure}

Audio signals encompass a rich variety of information, which broadly includes three main categories of elements: speech, audio/sound events and music. First, speech is distinguished by its linguistic content, which conveys semantic meaning, and also by paralinguistic information, including aspects such as emotion, accent, age, speaking style, intonation, and speaker identity. Second, environmental sounds or audio events refer to signals that offer insights into the surroundings, providing basic semantic meaning about the presence of an event or object, such as traffic noises or honking. While speech can inherently be part of sound events, the focus here is on the non-linguistic characteristics (\textit{e.g.} ``a man is speaking''). Third, music includes both singing (combining linguistic and musical elements) and pure instrumental music. Inherently, audio signals are time sequences that carry crucial temporal information, and in the real world, they also contain spatial information related to sound sources. Collectively, understanding these audio elements constitutes the foundational building blocks of a nascent concept of computer audition \cite{wang2010machine}, which aims to equip artificial systems with generic human-like auditory perception and cognition abilities.

% Empowering LLMs with audio modality has led to the development of audio LLMs (also refers to as auditory LLM, speech LLM, and spoken LLM), which demonstrate a wide array of capabilities across all audio-related tasks. These models excel at audio perception and understanding, performing tasks such as automatic speech recognition (ASR), speech translation, and spoken language understanding (SLU) across diverse languages, accents, and domains. Beyond basic transcription, they can engage in complex audio analysis, including audio event detection, music captioning, and audio (scene) understanding, discerning subtle relationships between auditory and visual signals to interpret environmental sounds and general audio events. Crucially, audio LLMs also possess significant generative abilities, capable of producing textual responses from audio inputs, synthesizing natural speech (both text-to-speech and speech-to-speech), and controlling paralinguistic aspects like emotion, style, timbre, and speaker identity. Their advanced multimodal understanding enables them to tackle open-ended and complex reasoning tasks, such as audio question answering (QA), and even temporal localization of sound sources within video, facilitating seamless and natural real-time human-computer interactions, including full-duplex conversations with dynamic turn-taking.

The advent of LLMs is fundamentally reshaping the paradigm of audio processing, propelling the field of audio and speech processing into a new era. We are moving beyond traditional tasks and are now striving for more comprehensive understanding, more natural generation, and more human-like interaction. This marks a decisive shift towards imbuing machines with genuine ``auditory intelligence". 

\begin{itemize}
    \item In the realm of auditory understanding, we are transcending the boundaries of traditional tasks like Automatic Speech Recognition (ASR), speaker verification, or sound event detection. The new paradigm demands a holistic perception that encompasses both temporal dimensions (e.g., the sequence and duration of events) and spatial dimensions (e.g., the movement and proximity of sound sources). This means a machine must not only ``hear" sounds but also ``comprehend" the physical world behind them, enabling complex reasoning and inference. For instance, from a single recording, a model should not only identify footsteps and a door closing but also infer the higher-level event that ``someone has left the room''. 
    \item In auditory generation, the ambition is to create sound that is indistinguishable from reality. This requires generative models to achieve exceptional naturalness, controllability, and stability. The frontier of this research includes synthesizing natural human voices with specific emotions, accents, and styles; creating complex and diverse soundscapes, from a bustling city street to a tranquil forest; and composing fluid and expressive music. The core objective is to move beyond mechanical concatenation and towards lifelike, dynamic creation.
    %\item In auditory interaction, the ultimate goal is to facilitate conversations as seamless and natural as those between humans. This necessitates models capable of handling real-time interruptions (barge-in), accurately tracking the dynamics of multi-person interactions, and effectively navigating acoustically complex scenarios like the cocktail party problem. Such interaction is not merely an exchange of information but a complete experience involving dynamic turn-taking, empathetic resonance, and instantaneous feedback. 
    \item In the domain of auditory interaction, the ultimate objective is to create a conversational experience that is as seamless and natural as human-to-human dialogue. This requires models to master conversational behaviors, including dynamic turn-taking, handling real-time interruptions (barge-in), and providing instantaneous feedback such as backchanneling. Beyond just reacting, a truly intelligent agent must also demonstrate proactivity. This involves models perceiving subtle cues from the acoustic environment and the user's voice to anticipate needs or changes. This proactive ability moves the interaction from a simple exchange of information to a more complete and empathetic experience.
    \item Finally, as a critical component of how humans perceive the world, audition is not destined to evolve in isolation in the age of AGI. It must develop in synergy with other senses, particularly vision. The recent proliferation of audio-visual models exemplifies this trend. By integrating information from both auditory and visual streams, these models can achieve a more robust and nuanced perception of the physical world, enabling them to tackle far more complex reasoning and interactive tasks and taking a firm step towards truly comprehensive machine intelligence.
\end{itemize}

The integration of the audio modality is essential for developing AGI, as human cognition inherently processes information by various senses, including sound, to understand and interact with the environment. While existing surveys have predominantly focused on speech modality \cite{ji2024wavchat,peng2024survey,cui2024recent}, and some have extended their scope to general audio \cite{arora2025landscape,su2025audio}, multi-modality integration, such as audio-visual integration, remains largely overlooked. Therefore, this survey aims to provide a comprehensive overview of the audio modality in LLMs, covering all LLM applications related to audio processing. This survey will delve into the multifaceted integration of audio modality within LLMs, structuring our analysis across several key dimensions. First, we explore audio representation as a background, examining how raw audio signals are effectively converted into representations consumable by LLMs, often involving discrete speech units or continuous embeddings. Second, we detail audio as input for understanding, focusing on architecture, training, applications, and evaluation. Third, the survey address audio as output for generation, covering the synthesis of speech, sound event, and music. Fourth, we examine the paradigm of speech interaction, emphasizing models designed for natural human-computer spoken dialogue, including multi-turn conversations and the nuanced handling of paralinguistic information. Finally, we dedicate a section to audio-visual integration, which explores how LLMs combine auditory and visual information for a comprehensive understanding of dynamic scenes and events.

\section{Audio Representation for LLMs}
\label{Sec:2}

\begin{figure}[ht]
    \centering
    \includegraphics[width=\linewidth]{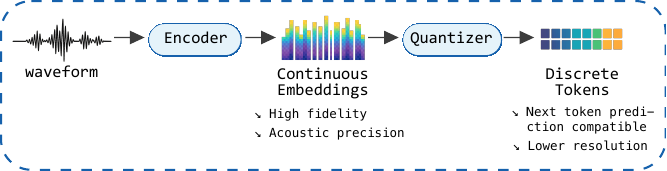}
    % \caption{Two paradigms for audio representation in LLMs: Continuous embeddings preserve high-fidelity acoustic details for comprehension, while discrete tokens offer token-based compatibility for generation tasks. Each path processes raw waveforms through different encoders—continuous encoders for dense embeddings and neural codecs for LLM-compatible discrete tokens—highlighting a core trade-off between fidelity and generative efficiency.}
    \caption{Two paradigms for audio representation in LLMs: Continuous embeddings preserve high-fidelity acoustic precision beneficial for comprehension tasks, whereas discrete tokens, generated via quantization, provide lower-resolution but token-compatible representations optimal for generative tasks like next-token prediction. Both paradigms encode raw waveforms differently, balancing fidelity with generative efficiency.}
    \label{fig:chpt2}
\end{figure}

The rapid advancements in LLMs have profoundly transformed natural language processing, enabling unprecedented capabilities in text comprehension and generation. However, extending these powerful models to multimodal domains, particularly audio, introduces a significant and inherent challenge: the intrinsic mismatch between audio's continuous signal structure and the discrete, token-based architecture of LLMs. To bridge this critical modality gap, researchers have primarily explored two distinct paradigms for representing audio for LLMs: continuous embeddings and discrete tokens.

\subsection{Continuous embeddings}

Continuous audio embeddings represent audio signals as dense, high-dimensional vectors, designed to preserve a rich array of acoustic details without explicit quantization. These representations are typically learned through either self-supervised pre-training on vast datasets of unlabeled audio, such as HuBERT \cite{hubert} and BEATs \cite{beats}, or large-scale supervised training on specific downstream tasks, as seen in models like Whisper \cite{whisper} and USM \cite{usm}. An audio encoder processes raw waveforms or Mel spectrograms to produce these continuous embeddings, which are often downsampled and projected into representations compatible with the LLM's input space. The core principle behind continuous audio embeddings is to maximize the preservation of information from the original audio signal. This makes them particularly well-suited for tasks requiring fine-grained acoustic distinctions or high-resolution audio comprehension, provided that the subsequent LLM architecture can effectively process such dense and high-dimensional input. Consequently, many audio-focused LLMs \cite{salmonn,qwenaudio,qwen2audio,wavllm,gama,audioflamingo} adopt continuous embeddings to enhance audio understanding performance. However, despite their advantages in comprehension tasks, continuous representations pose substantial challenges for audio generation. Their non-discrete nature conflicts with the autoregressive, token-by-token generation paradigm typically used in LLMs, which is inherently designed for discrete token spaces. While emerging approaches such as E2TTS, MELLE, and DiTAR \cite{e2tts,melle,ditar} explore speech generation from continuous embeddings, effectively leveraging LLMs for this purpose remains an open research question. Further work is needed to develop robust strategies for generating speech and audio from such continuous latent spaces.

\subsection{Discrete tokens}

Discrete audio tokenization is a paradigm that transforms continuous audio signals into sequences of discrete, quantized units, thereby directly mirroring the token-based input format of LLMs. This process typically involves neural audio codecs or advanced vector quantization techniques. The core principle is to discretize continuous audio features to better align with the token-based paradigm of LLMs. Neural codecs \cite{soundstream,encodec,dac,casanova2025nanocodec}, which are central to this approach, produce discrete audio tokens with the discretization performed by a differentiable quantizer, such as residual vector quantization (RVQ) and finite scale quantization (FSQ) \cite{fsq}. Discrete audio tokens can also be derived by applying clustering algorithms, such as k-means, to the continuous embeddings of pre-trained encoders like HuBert \cite{hubert} and wav2vec 2.0 \cite{w2v2}. These tokens are commonly adopted by ``textless'' natural language processing (NLP) studies, such as GSLM \cite{gslm} and dGSLM \cite{dGSLM}, which achieve NLP tasks relying only on speech rather than texts. In contrast to continuous embeddings, which emphasize preserving fine-grained acoustic detail, recent research on discrete audio tokens has increasingly focused on compression and disentanglement to enable more efficient and interpretable modeling. Compression aims to shorten audio token sequences to lengths comparable to text, making them easier for LLMs to model. Notable work in this direction includes SingleCodec \cite{singlecodec} and WavTokenizer \cite{wavtokenizer}, which push the limits of temporal compression. On the other hand, disentanglement focuses on extracting semantic tokens that are more readily interpretable by text-based LLMs. To achieve this, models like SpeechTokenizer \cite{speechtokenizer} and NaturalSpeech 3 \cite{naturalspeech3} incorporate semantic distillation into codec training, while CosyVoice \cite{cosyvoice} directly leverages ASR-driven, semantic-centric tasks to guide the construction of discrete representations.

\subsection{Discussion}

The choice between continuous embeddings and discrete tokens for audio representation in LLMs requires a careful evaluation of their respective advantages and trade-offs. Continuous embeddings excel in fidelity, preserving the maximum amount of acoustic information, including subtle nuances and fine-grained details \cite{wang2025speech}. This high fidelity is crucial for tasks that require precise acoustic distinctions. Discrete tokens prioritize compatibility and efficiency. By converting audio into a discrete format, they become structurally analogous to text tokens, allowing for seamless integration into existing LLM architectures, and naturally support autoregressive generation through next-token prediction. While powerful modality-specific encoders already exist, such as WavLM \cite{wavlm} for speech, BEATs \cite{beats} for general audio, and MERT \cite{mert} for music, a truly unified encoder capable of handling diverse audio modalities remains an open challenge. Despite pioneering efforts such as MT2KD \cite{mt2kd} and Dasheng \cite{dinkel2024dasheng,dinkel2025midashenglm}, general-purpose audio modeling still lacks a standardized solution. Recently, challenges \cite{codecsuperb,icmechallenge} are also promoting robust benchmarks and fostering the development of more powerful and versatile audio encoders. Looking ahead, the overarching objective for audio representation in LLMs is clear: more effective and more efficient.

The central bottleneck in current codec research arises from the tension between semantic clarity and paralinguistic fidelity: continuous embeddings offer acoustic precision at the cost of token compatibility, while discrete tokens trade subtle acoustic detail for seamless LLM integration. Recent approaches, including hierarchical tokenization, adaptive bitrate allocation, and improved quantizers, aim to efficiently reconcile these trade-offs. Standardized benchmarks further guide the community towards codecs that balance semantic robustness and acoustic expressiveness within practical computational limits.

\section{LLMs for Audio Comprehension}

\begin{figure}[ht]
    \centering
    \includegraphics[width=\linewidth]{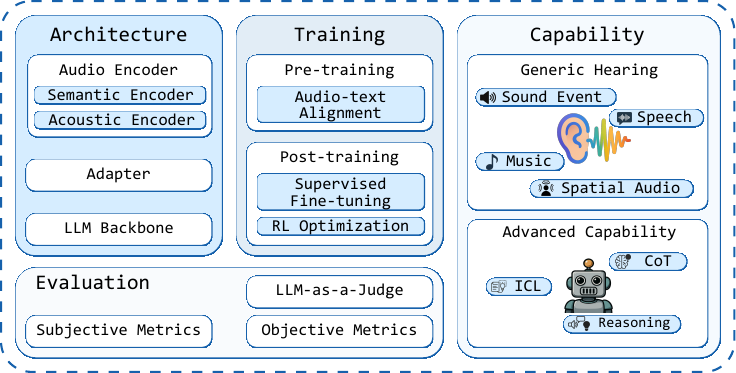}
    \caption{An overview of Audio LLMs: from architecture (encoder, adapter, LLM) and modality-aligned training to human-like comprehension, in-context learning, and reasoning across diverse audio modalities.}
    \label{fig:chpt3}
\end{figure}

\begin{figure*}[ht]
    \centering
    \includegraphics[width=0.8\linewidth]{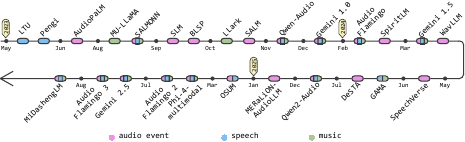}
    \caption{Timeline of understanding-centric audio LLMs, where different colors indicate the categories of audio signals each model can process. Some works not mentioned in the main text are cited here \cite{abouelenin2025phi,ghosh2025audioflamingo2,goel2025audioflamingo3}}
    \label{fig:allm_timeline}
\end{figure*}

Audio LLMs represent a burgeoning field dedicated to achieving universal understanding and complex reasoning across diverse audio elements, including speech, music, audio events, the starting and ending time, and the location of the sound source. 
%general environmental sounds. 
The profound significance of audio LLMs lies in their capacity to move beyond mere sensory-level tasks, such as simple transcription or classification, to engage in complex cognitive processes that mirror human auditory comprehension. In this section, the typical architecture, training paradigms, applications and evaluation of audio LLMs for audio comprehension will be discussed.

\subsection{Architecture}

The typical architectural design of audio LLMs fundamentally involves three components: an audio encoder, a modality adapter (often termed a connector or projector), and an LLM backbone. This framework facilitates the integration of complex auditory signals with the linguistic processing capabilities of LLMs.

The initial component, the audio encoder, serves the crucial function of converting raw waveforms into representations that are compatible with the language model. This encoding can manifest in two primary forms: continuous embeddings or discrete tokens, as discussed in Section~\ref{Sec:2}. Most dedicated encoder designs focus on enhancing the model's ability to process general audio signals effectively. Given the absence of a powerful universal audio encoder, models targeting comprehensive auditory understanding across diverse sound types, such as SALMONN \cite{salmonn} and WavLLM \cite{wavllm}, often adopt a dual-encoder architecture, integrating both a speech encoder (e.g., Whisper \cite{whisper}) and an acoustic encoder (e.g., BEATs \cite{beats} or WavLM \cite{wavlm}). An alternative strategy is employed by Prompt-Aware Mixture of Experts \cite{shan2025enhancing}, which dynamically selects specialized encoders based on the input prompt to extract task-relevant features for improved generalization. Other multi-task audio LLMs, including Qwen-Audio \cite{qwenaudio} and Qwen2-Audio \cite{qwen2audio}, opt to fully fine-tune pre-trained speech encoders on diverse audio understanding tasks, enabling the encoder to adapt and generalize beyond speech. Similarly, SOLLA \cite{ao2025solla} augments the encoder with an audio tagging module to enhance audio information extraction during fine-tuning. In the vision domain, there are also efforts to build vision LLMs without an encoder \cite{diao2024unveiling,luo2025mono}, which have not yet been extended to the audio modality.

The modality adapter, often referred to as a projector or connector, plays a pivotal role in aligning the audio encoder's output with the LLM backbone. It acts as a bridge, transforming speech representations into the latent embedding space expected by the language model. For discrete tokens, this modality adapter is typically a simple embedding layer. In contrast, for continuous embeddings, more complex architectures are employed, such as multi-layer perceptrons (MLPs) \cite{qwenaudio,qwen2audio,he2024meralion}, window-based Q-Formers \cite{salmonn}, or Conformer-based modules \cite{chen2024salm}. The effectiveness of different connector designs has been systematically evaluated in recent studies \cite{yu2024connecting,ma2025speech}, which suggest that the optimal choice may vary depending on the dataset and task. To enhance audio comprehension, GAMA \cite{gama} introduces multiple parallel connectors, enabling the integration of diverse audio features for improved understanding. Since the core function of the modality adapter is to align speech representations with the LLM's internal structure, one key research direction is improving this alignment. Techniques such as CTC-based \cite{fan2024alignformer,zhang2025soundwave} and CIF-based \cite{deng2025wav2prompt} compression have shown promise in achieving tighter modality matching between speech and text, leading to improved instruction-following performance in speech-based tasks. However, these alignment methods are currently tailored to speech and cannot be directly applied to general audio processing, where the lack of linguistic structure poses additional challenges.

The LLM backbone, serving as the central sequence modeling component, is typically built upon a pre-trained text language model from leading LLM families such as LLaMA \cite{touvron2023llama}, Qwen \cite{bai2023qwen}, or T5 \cite{raffel2020exploring}. This backbone is responsible for processing the fused audio-text representations and generating the final outputs. While most LLM backbones retain a decoder-only architecture \cite{ltu,audiopalm,salmonn,qwen2audio}, the Flamingo-style cross-attention mechanism has also been explored in \cite{li2023prompting,audioflamingo}. To preserve the original language modeling capabilities and reduce catastrophic forgetting when incorporating audio inputs, the LLM backbone is typically kept frozen \cite{qwenaudio,qwen2audio} or fine-tuned using lightweight adaptation techniques such as LoRA \cite{salmonn,gama}. WavLLM \cite{wavllm} employs a prompt-aware LoRA weight adapter for optimized performance. While decoder-only Transformers remain the dominant architecture, emerging alternatives like state-space models, particularly Mamba \cite{gu2023mamba}, and diffusion LLMs \cite{zhou2025diffa} are gaining attention for their potential to improve efficiency in long-context modeling \cite{bhati2024state,lu2025duplexmamba}.

\subsection{Training}

The training stages of text LLMs are commonly classified into two stages: \textit{pre-training} and \textit{post-training}. During \textit{Pre-training}, the model is exposed to vast general-purpose corpora to build a broad understanding of language. \textit{Post-training} then refines the model for specific tasks, enhancing its target capabilities, accuracy, and alignment with user intent. A similar two-stage framework can be applied to audio LLMs. In this context, \textit{pre-training} focuses on integrating the audio modality into a pre-trained text LLM, laying the groundwork for basic auditory perception. \textit{Post-training} then further refines the model for task-specific performance and specialized capabilities. It is worth noting that Gemini \cite{team2023gemini}  and GPT-4o explore native multimodal pre-training, which combines text pre-training and audio pre-training into a unified pre-training stage. This unified approach leads to improved multimodal understanding. However, such large-scale training requires significant data and computational resources, which are often beyond the reach of academic institutions.

\textit{Pre-training} stage of audio LLMs focuses on modality adaptation and alignment, bridging the intrinsic gap between speech and text modalities. The primary objective here is to align information from different modalities into a unified embedding space. Pre-training typically employs supervised fine-tuning (SFT) as the training method. Since the modality adapters (connectors) are usually randomly initialized, curriculum learning is widely adopted \cite{salmonn,qwenaudio,wavllm,gama,audioflamingo,das2024speechverse} to ensure stable training. A common strategy begins with aligning the audio modality using simpler tasks, followed by instruction tuning on more complex and diverse datasets. For instance, LTU \cite{ltu} trains the connector using only simple classification and description tasks at the first pre-training stage, then trains the audio encoder, the connector and LoRA on LLM using all data. Although ASR is commonly used in first-stage pre-training, recent studies show that better alignment can be achieved through continuation-based tasks \cite{audiochatllama,wang2023blsp} or self-powered data \cite{yu2024self}. The components involved in training are highly flexible. Many models adopt LoRA-based adapters on the LLM while freezing both the audio encoder and the LLM backbone, training only the connector and LoRA modules \cite{salmonn,wavllm,gama,das2024speechverse}. Others opt to freeze the LLM entirely and instead train the audio encoder and connector to bridge the modality gap \cite{qwenaudio,pengi}. In addition, some approaches use knowledge distillation for pre-training \cite{held2024distilling}, enabling the model to achieve strong instruction-following performance without explicit instruction tuning. Recently, interleaved pre-training \cite{spiritlm,glm4voice,kimiaudio,wang2025inserter} is a new fashion to mitigate the modality gap, which follows the paradigm of text pre-training with portions of text input substituted by speech, resulting in better preserved text abilities of LLM backbone such as semantic information understanding and instruction following. 

\textit{Post-training} plays a critical role in enhancing an audio LLM's performance on specific tasks and advancing its capabilities in areas such as in-context learning (ICL) and reasoning. Two primary training methods are commonly employed: SFT and reinforcement learning (RL). SFT typically involves curating specialized datasets and fine-tuning the audio LLM on these tasks to equip it with new skills or strengthen complex capabilities. For example, \cite{wang2025enabling} and \cite{chen2025audio} fine-tune audio LLMs to evaluate speech quality through an LLM-as-a-judge framework. Similarly, MALLM \cite{chen2024beyond} is trained to distinguish between speech pairs, thereby improving its ability to process multiple audio inputs. Reinforcement learning, particularly through methods such as PPO \cite{ppo}, DPO \cite{dpo}, and GRPO \cite{grpo}, has emerged as a powerful technique for fine-grained performance optimization. It is especially effective at enhancing reasoning abilities and aligning model outputs with human preferences. For instance, Qwen2-Audio \cite{qwen2audio} and Seed-ASR \cite{seedasr} utilize DPO to improve factual accuracy and behavioral alignment. More recently, several works \cite{li2025reinforcement,rouditchenko2025omni,wijngaard2025audsemthinker,wu2025audio} have leveraged GRPO to develop reasoning-capable audio LLMs that can ``think before answering'', leading to a boosted performance.

To build a custom audio LLM, researchers can start with speech processing toolkits such as ESPnet \cite{watanabe2018espnet} or SLAM-LLM \cite{ma2024embarrassingly}. Alternatively, they can adapt the fine-tuning scripts provided by popular open-source audio LLMs like SALMONN, Qwen2.5-Omni, and Kimi-Audio. In addition, ongoing efforts within academia are pushing for greater openness and transparency in audio LLM development, as demonstrated by projects such as OSUM \cite{geng2025osum} and OPUSLM \cite{tian2025opuslm}.

\subsection{Capability}

Audio LLMs are demonstrating remarkable versatility, addressing an expansive spectrum of tasks that encompass text, speech, music, and general audio functionalities. Their application domains are rapidly expanding beyond conventional audio processing.

In the realm of speech-related applications, audio LLMs have significantly advanced capabilities in semantic tasks such as Automatic Speech Recognition (ASR) \cite{li2023prompting,chen2023x}, Speech-to-Text Translation (S2TT) \cite{chen2024salm,wu2023decoder}, Spoken Question Answering (SQA) \cite{nachmani2023spoken}, Spoken Language Understanding (SLU) \cite{shon2024discreteslu} and spoken dialogue \cite{lin2024paralinguistics,xue2024chat}. Beyond these, they also contribute to speaker-related applications such as speaker identification \cite{wu2024just}, speaker verification \cite{salmonn}, and speaker diarization \cite{team2023gemini}. A notable expansion includes other paralinguistic applications, enabling emotion recognition \cite{wu2024beyond}, accent recognition \cite{mu2024mmger,jairam2024few}, gender recognition \cite{wang2025qualispeech}, speech quality assessment \cite{wang2025enabling,chen2025audio}, spatial speech understanding \cite{zheng2024bat,tang2024can} and speaking style recognition \cite{lin2024advancing}. As for the sound and music domain, applications involve Automatic Audio Captioning (AAC) \cite{pengi,tang2024extending}, Audio Question Answering (AQA) \cite{ltu,deshmukh2025audio} and music question answering and captioning \cite{doh2023lp,gardner2023llark,liu2024music}. 

Next, we further discuss the advanced capabilities of audio LLMs, which are critical for expanding their applicability across diverse, real-world scenarios. One such key ability is instruction following, which allows models to generalize to unseen tasks by interpreting natural language prompts. However, after pre-training integrating audio modality, audio LLMs often suffer from catastrophic forgetting, resulting in poor generalization to novel instructions. To mitigate this, several techniques have been proposed, including CTC-based alignment \cite{fan2024alignformer}, activation tuning \cite{salmonn}, and the incorporation of text modality supervision \cite{lu2024desta}. An interesting phenomenon observed in \cite{wang2025textbias} is that audio LLMs display a significant bias toward textual input when audio and text disagree, suggesting the audio modality alignment is fragile. Dealing with this problem may provide a new perspective for better instruction following.
Another essential skill is in-context learning (ICL), which enables audio LLMs to quickly adapt to new tasks using only a few examples provided at inference time. This ability has been actively explored in recent works \cite{audioflamingo,pan24b_interspeech,liang2025acoustic}. Nevertheless, due to the limited availability of high-quality, multi-audio paired datasets, current open-source audio LLMs still struggle to achieve strong ICL performance on truly novel tasks.

An especially promising and emerging capability is reasoning, where models learn to ``think before answering'', often yielding substantial gains through test-time scaling. Test-time Chain-of-thought (CoT) prompting for audio LLMs has been shown to be effective in recent studies \cite{kuan2025can,ma2025audio}, where models are instructed to first generate audio descriptions before answering specific audio-related questions, leading to measurable gains in accuracy. Moreover, audio-speech co-reasoning, as investigated in \cite{salmonn,ao2025solla}, involves analyzing and synthesizing multiple facets of the audio signal to support a more holistic understanding. Inspired by models like OpenAI-o1 \cite{jaech2024openai} and DeepSeek-R1 \cite{guo2025deepseek}, several recent audio LLMs have been developed to explicitly incorporate reasoning capabilities for improved comprehension \cite{xie2025audio,deshmukh2025mellow,li2025reinforcement,rouditchenko2025omni,wijngaard2025audsemthinker,wu2025audio}. These advancements collectively underscore the growing potential of audio LLMs to revolutionize human-computer interaction, particularly in complex, real-world auditory scenarios.

\subsection{Evaluation}

Evaluating audio LLMs comprehensively and fairly remains a multifaceted challenge, largely due to their wide-ranging capabilities and the fragmented nature of existing benchmarks. A number of general-purpose benchmarks have been developed to provide broad coverage and holistic evaluation of audio comprehension, including Dynamic-SUPERB \cite{huang2024dynamic} and its extended Phase-2 version \cite{huang2024dynamic2}, AIR-Bench \cite{yang2024air}, AudioBench \cite{wang2024audiobench}, SAGI \cite{bu2024roadmap}, MMAU \cite{sakshi2024mmau} SALMON \cite{maimon2025salmon}, MMSU \cite{wang2025mmsu} and MMAU-Pro \cite{kumar2025mmau}. These benchmarks aim to assess audio LLMs across a diverse set of tasks and scenarios. For example, Dynamic-SUPERB Phase-2 includes an extensive suite of 180 audio-related tasks, reflecting the breadth of the modality.

Beyond general benchmarks, task-specific benchmarks target nuanced aspects of audio comprehension. In paralinguistic understanding, datasets like SD-Eval \cite{ao2024sd}, StyleTalk \cite{lin2024advancing}, VoxDialogue \cite{cheng2025voxdialogue}, and E-chat200 \cite{xue2024chat} assess a model's ability to interpret nuanced cues such as emotion, accent, age, and speaking style. QualiSpeech \cite{wang2025qualispeech} evaluates low-level speech perception, while Finaudio \cite{cao2025finaudio} focuses on financial audio comprehension. In the music domain, benchmarks such as MuChoMusic \cite{weck2024muchomusic}, OpenMU-Bench \cite{zhao2024openmu}, and CMI-Bench \cite{ma2025cmi} test models on musical understanding. Broader concerns related to fairness and safety are addressed in benchmarks examining semantic gender bias \cite{lin2024listen}, trustworthiness \cite{li2025audiotrust}, and jailbreak vulnerabilities, as explored by JailBreak-AudioBench \cite{cheng2025jailbreak}, JALMBench \cite{peng2025jalmbench} and WhisperInject \cite{kim2025good}. The issue of hallucination, where models generate plausible but incorrect audio-related outputs, is examined in \cite{kuan2024understanding}. For advanced capabilities, Speech-IFEval \cite{lu2025speech} evaluates instruction-following and specifically targets the problem of catastrophic forgetting. MAE-Bench \cite{chen2024beyond} focuses on multi-audio processing, a core requirement for effective ICL. Long-form audio comprehension, essential for realistic human interaction, is measured by BLAB \cite{ahia2025blab}. Reasoning abilities are rigorously tested in MMAR \cite{ma2025mmar} and SAKURA \cite{yang2025sakura}, which assess performance on multi-step logical inference. Furthermore, JASCO \cite{wang2025they} investigates audio-speech co-reasoning, emphasizing joint analysis of linguistic and acoustic information.

Evaluation methodologies typically fall into two main categories: automatic (objective) metrics and human assessments. Automatic evaluation relies on well-established metrics such as word error rate (WER) for speech recognition, accuracy for question answering, and text generation scores like BLEU, METEOR, and ROUGE for tasks such as translation and summarization. 
While human evaluations remain essential for assessing subjective qualities, including but not limited to naturalness, emotional expressiveness, and instructional clarity, they are often costly and time-consuming. To address this bottleneck, LLM-as-a-judge methods \cite{zheng2023judging} are increasingly being adopted. These approaches simulate human evaluation using language models and have demonstrated strong correlation with human ratings, offering a scalable and efficient alternative for evaluating open-ended outputs. Despite these advances, several critical challenges persist. Key challenges include issues such as data contamination and insufficient consideration of human diversity within existing datasets. The limited diversity and scale of audio data sources further impede robust training and evaluation. Quality issues in synthetically generated datasets, particularly concerning inaccuracies and hallucinations, remain a concern.

\section{LLMs with Audio as Output for Generation}

\begin{figure}[ht]
    \centering
    \includegraphics[width=\linewidth]{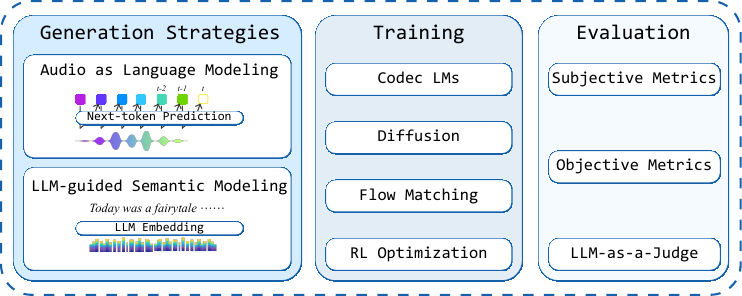}
    \caption{An overview of LLM-based audio generation: combining token-based and semantic modeling strategies, hybrid training methods, and human-aligned evaluation using both objective metrics and LLM-as-a-judge frameworks.}
    \label{fig:chpt4}
\end{figure}

In the evolving landscape of artificial intelligence, the integration of LLMs with auditory modality has emerged as a pivotal area, leading to advanced capabilities in audio generation. This section details the current state of audio LLMs for audio generation, categorizing key strategies, training methodologies, and evaluation approaches.

\subsection{Generation Strategies}

\subsubsection{Audio generation as language modeling}

The application of language modeling techniques to audio generation, where discrete audio tokens are generated autoregressively using Transformer-based architectures, was pioneered in 2022 by studies such as AudioLM \cite{borsos2023audiolm} and VALL-E \cite{wang2023neural}. These early models demonstrated impressive performance, particularly in zero-shot text-to-speech (TTS) synthesis, establishing a strong foundation for subsequent research. It is worth highlighting that language modeling techniques have also achieved notable success in text-to-music generation \cite{agostinelli2023musiclm,copet2023simple} and video-to-audio generation \cite{kondratyuk2023videopoet}.
Building on this progress, researchers have increasingly pursued the idea of a universal speech generation model, inspired by the multi-task generalization capabilities of text-based LLMs. Models like SpeechX \cite{wang2024speechx}, VioLA \cite{wang2024viola}, and UniAudio \cite{yang2023uniaudio} aim to develop versatile audio generation models capable of handling a broad spectrum of speech and audio transformation tasks. These include zero-shot TTS, noise suppression, speech enhancement, speech editing, target speaker extraction, and speech separation. pre-trained on large-scale datasets covering diverse tasks, these models demonstrate strong generalization and can be readily fine-tuned for new downstream applications, highlighting their potential as foundation models for audio generation.

\subsubsection{Leveraging LLM for semantic modeling}

A growing trend in audio generation involves leveraging the advanced semantic understanding of LLMs to enhance the quality, coherence, and expressiveness of synthesized speech and general audio. Besides approaches that enhance the video-to-audio pipeline by leveraging captioning capabilities of LLMs \cite{xie2024sonicvisionlm,liu2024tell}, we mainly focus on end-to-end audio generation with LLM.

In speech generation, models such as CosyVoice2 \cite{du2024cosyvoice2} integrate a Qwen2.5-0.5B LLM for text-to-token conversion, followed by a conditional flow-matching model for token-to-speech synthesis. Muyan-TTS \cite{li2025muyan} pairs a LLaMA-3.2-3B LLM with a VITS-based decoder \cite{kim2021conditional}, aligning text and audio through quantized acoustic tokens. GOAT-TTS \cite{song2025goat} mitigates catastrophic forgetting by freezing lower layers of the LLM, preserving its innate language understanding. VibeVoice \cite{peng2025vibevoice} streamlines speech synthesis by concatenating text and voice features as input to an LLM, whose hidden states condition a lightweight diffusion head for continuous token prediction, enabling scalable long-form multi-speaker generation. Other approaches, such as \cite{hao2025boosting}, explore directly fine-tuning text LLMs as codec language models, akin to VALL-E. Collectively, these methods have achieved notable gains in intelligibility, naturalness, and speaker similarity, underscoring the benefits of incorporating LLMs into speech generation. 

In the broader context of general audio generation, TANGO \cite{deepanway2023text} employs an instruction-tuned LLM (FLAN-T5) \cite{chung2024scaling} as the text encoder, combined with a latent diffusion model for synthesis. Despite being trained on a relatively small dataset, TANGO surpasses previous SOTA text-to-audio systems, attributing its success to FLAN-T5's strong representational power derived from instruction tuning.  Make Some Noise (LM-MSN) \cite{mehta2025make} explores unifying audio comprehension and generation within a single LLM framework. While LoRA-based fine-tuning of a pre-trained text LLM showed promise for comprehension, generation quality remained suboptimal, revealing a critical need for larger, more diverse training data.

\subsection{Training}

Training objectives in audio generation models vary based on architectural design. For codec-based language models \cite{borsos2023audiolm,wang2023neural}, the cross-entropy loss is employed to optimize next-token prediction. Diffusion model \cite{ho2020denoising} is trained by minimizing the expected mean squared error between the noise it predicts and the actual Gaussian noise. Flow matching approaches \cite{lipman2022flow} also rely on MSE loss, but focus on aligning the model's predicted time-dependent vector fields with the ground truth conditional vector fields. RL methods have been explored to enhance the robustness of audio generation \cite{chen2024enhancing} and improve alignment with human preferences, as demonstrated by recent work \cite{anastassiou2024seed,zhang2024speechalign}.

Data scale remains a central driver in the development of powerful audio LLMs. Leading models typically rely on large-scale, high-quality datasets, with many industry players leveraging proprietary in-house data. In the open-source domain, widely used datasets include LibriSpeech (960 hours) \cite{panayotov2015librispeech}, LibriHeavy (50,000 hours) \cite{kang2024libriheavy}, and the English subset of Multilingual LibriSpeech (MLS) (44,500 hours) \cite{pratap2020mls}. Some approaches, like UniAudio, scale training data to an impressive 150,000 hours, while Kimi-Audio curates over 13 million hours of diverse audio data for its training. Muyan-TTS, specifically designed for podcast scenarios, relies on over 150,000 hours of raw speech data, emphasizing the importance of high-quality, task-specific datasets.

\subsection{Evaluation}

Evaluating audio LLMs capable of generating speech or general audio requires comprehensive and diverse benchmarks to capture performance across a wide range of tasks and modalities. For speech synthesis, commonly used evaluation datasets include LibriSpeech \cite{panayotov2015librispeech}, VCTK \cite{veaux2019vctk}, and the Seed-TTS test set \cite{anastassiou2024seed}. These datasets support evaluations of tasks such as zero-shot TTS and voice conversion. Beyond speech, the assessment extends to general audio, including sound events and music, with benchmarks such as MusicCaps \cite{agostinelli2023musiclm} for text-to-music synthesis and AudioCaps \cite{kim2019audiocaps} or Clotho \cite{drossos2020clotho} for text-to-sound generation, as seen in evaluations of UniAudio and LM-MSN.

To thoroughly evaluate both audio and speech generation, researchers rely on a mix of objective and subjective metrics. Objective measures often include the WER, which quantifies the accuracy of synthesized speech against its target transcription, and Speaker Similarity (SS), typically measured by automatic speaker verification models like WavLM, to assess the preservation of speaker identity. Perceptual metrics such as PESQ (Perceptual Evaluation of Speech Quality) \cite{rix2001perceptual} and DNSMOS (Deep Noise Suppression Mean Opinion Score) \cite{reddy2021dnsmos} are utilized to gauge speech quality, although their direct applicability to generative models may be limited due to processing artefacts not accurately captured by these signal-level scores. For general audio generation, Fréchet Audio Distance (FAD) \cite{kilgour2018fr} and KL divergence are employed to measure the similarity between generated and real audio distributions and to assess semantic retention, respectively. The Mel Cepstral Distortion (MCD) metric serves to evaluate spectral differences, particularly in tasks like speech removal. Despite the utility of these objective metrics, Mean Opinion Score (MOS) and MUSHRA (MUlti Stimulus with Hidden Reference and Anchor) \cite{series2014method} remain indispensable for subjective human assessment of perceived naturalness, overall quality, and the subtleties of expressive generation, as human perception is the ultimate arbiter of audio quality. Recently, a new trend has emerged: leveraging Audio-LLMs as evaluators to generate natural language quality assessments of speech outputs \cite{wang2025enabling, chen2025audio, chiang2025audio}. These models can provide nuanced, interpretable feedback that complements traditional metrics and offers scalable alternatives to human listening tests.

\section{Speech Interaction with Audio LLMs}

\begin{figure}[h]
    \centering
    \includegraphics[width=\linewidth]{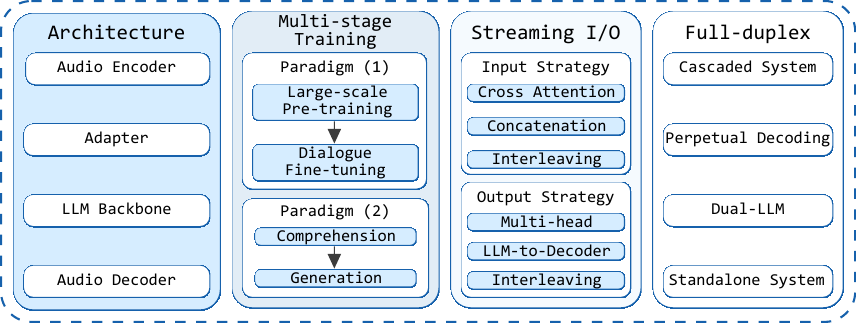}
    \caption{An overview of speech interaction LLMs: combining audio encoders, adapters, and decoders with training strategies, streaming I/O, full-duplex strategies and objective, subjective, and LLM-based metrics.}
    \label{fig:chpt5}
\end{figure}

\begin{figure*}[h]
    \centering
    \includegraphics[width=\linewidth]{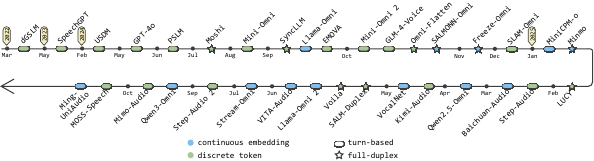}
    \caption{Timeline of end-to-end speech interaction LLMs. Models are categorized by generation strategies (utilizing LLM's continuous embedding or generating discrete tokens by LLM itself) and interaction mode (turn-based or full-duplex). Some works not mentioned in the main text are cited here \cite{gao2025lucy,huang2025step,shi2025voila,zhang2025stream,wu2025step,xu2025qwen3omni,coreteam2025mimoaudio,zhao2025moss,minguniaudio}}
    \label{fig:full_duplex_timeline}
\end{figure*}

\begin{figure*}[!tp]
\centering
\includegraphics[width=0.85\textwidth]{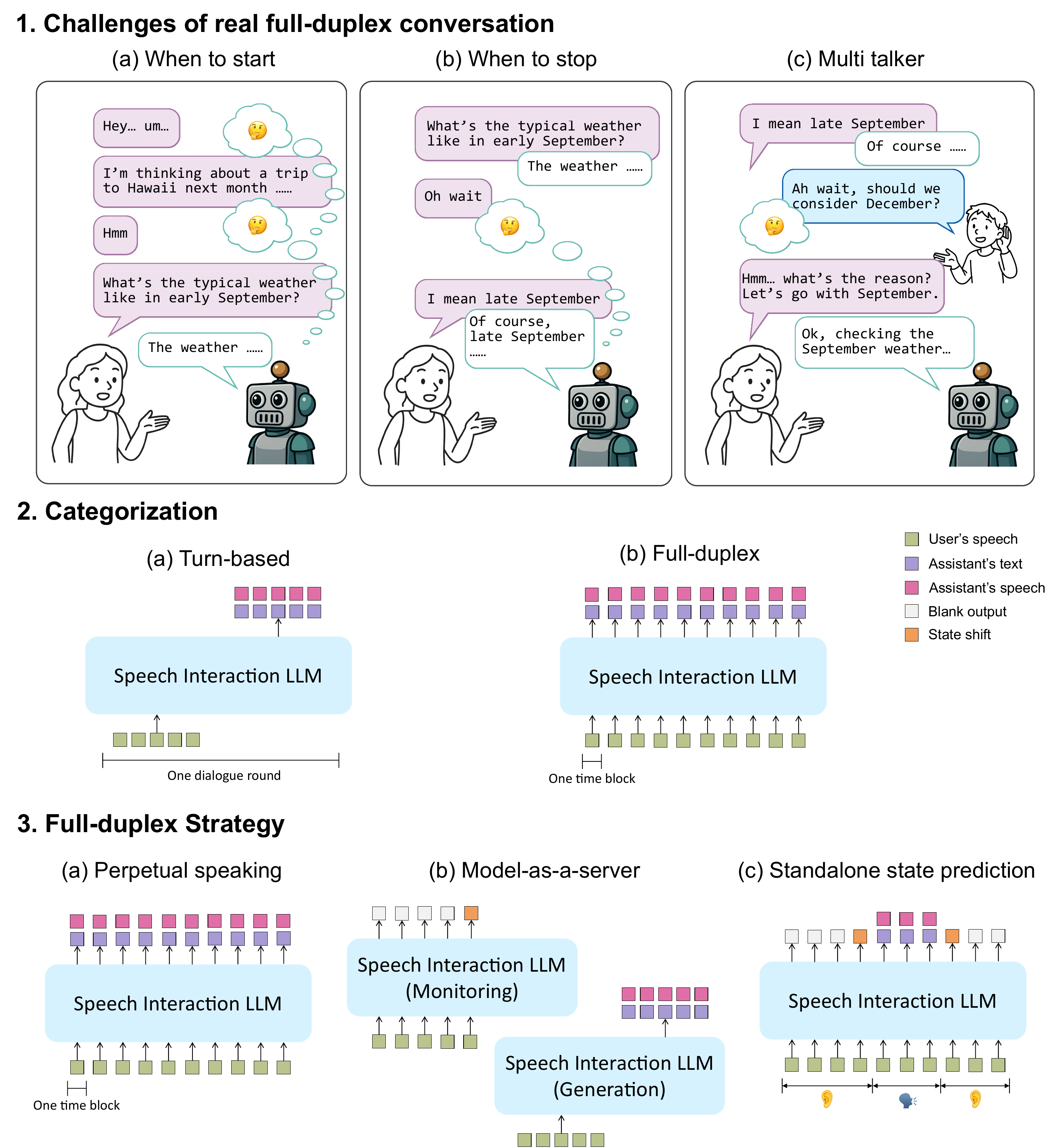}
\caption{Overview of speech interaction LLM architectures for full-duplex dialogue. The figure summarizes two key design dimensions: (1) Challenges of real full-duplex conversation, including when to start, when to stop and multi-talker scenario, (2) Interaction mode, contrasting turn-based and full-duplex systems and (3) Full-duplex implementation strategies, including perpetual speaking, model-as-a-server, and standalone state prediction.}
\label{fig:duplex_1}
\end{figure*}

Speech interaction LLMs, which unify speech understanding and generation, represent a significant leap in human-computer communication. They move beyond conventional text-based interfaces to embrace the richer, more complex modality of spoken language. However, achieving natural spoken dialogue, an indispensable capability for AGI, is far more challenging than the turn-based interactions that text LLMs excel at. While a text LLM can be readily adapted for a web interface, a true spoken dialogue system must be architected from the ground up to handle the fluid nature of human conversation. Full-duplex speech interaction models have attracted widespread attention following the release of GPT-4o (``o" for omni). Since then, many subsequent works have adopted ``omni" in their titles, such as LLaMA-Omni \cite{fang2025llamaomni} and Mini-Omni \cite{xie2024miniomni}, to emphasize their ability to support end-to-end speech interaction. This usage is recognized by the research community, however, it is noted that the prefix omni literally means ``all''. In this broader sense, models capable of perceiving multiple modalities (text, audio, images, and video) while simultaneously generating both text and speech in a streaming manner, such as MiniCPM-o \cite{minicpm} and Qwen2.5-Omni \cite{xu2025qwen2}, more closely align with the context of ``omni" in GPT-4o. 

\begin{figure*}[!tp]
\centering
\includegraphics[width=0.8\textwidth]{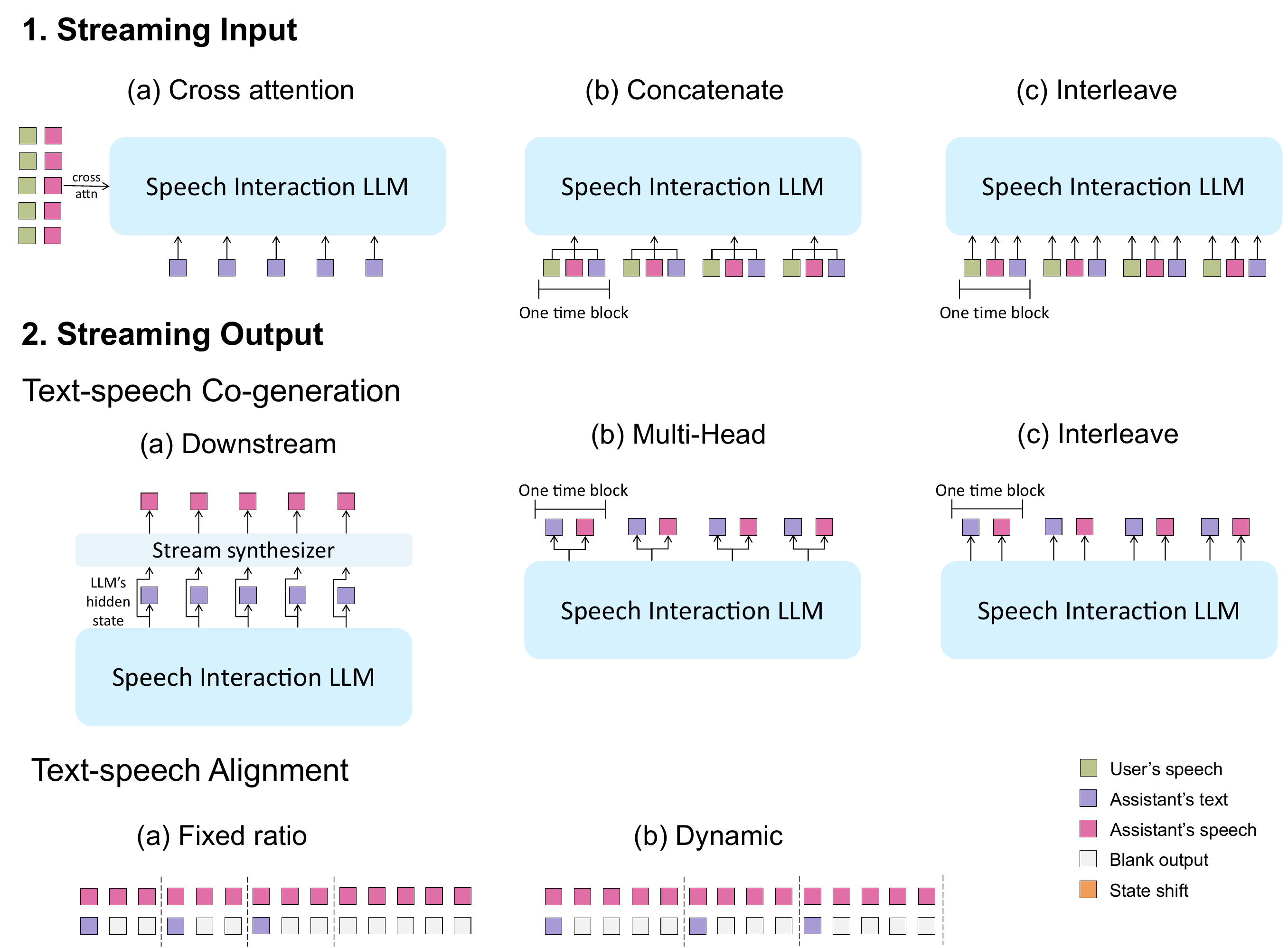}
\caption{Overview of streaming speech processing with LLM, which is the foundation for full-duplex dialogue. The figure contains two key aspects: (1) Streaming input strategies, including cross-attention, concatenation, and interleaving and (2) Streaming output strategies, such as downstream processing, multi-head co-generation, and interleaved decoding.}
\label{fig:duplex_2}
\end{figure*}

Real-world conversations present numerous complex challenges, as illustrated in Figure \ref{fig:duplex_1}. A model must learn to navigate intricate conversational dynamics such as turn-taking (knowing when to speak and when to yield), handling interruptions (barge-in), and providing subtle vocal feedback (backchanneling). Moreover, practical applications must contend with multi-speaker environments and solve the classic ``cocktail party problem''. To address these challenges effectively, the system must operate in a full-duplex manner, processing audio input while simultaneously generating a spoken response. This requirement for seamless, simultaneous interaction necessitates advanced streaming designs, as detailed in Figure \ref{fig:duplex_2}. Given this unique fusion of understanding and generation and the profound architectural demands of full-duplex interaction, end-to-end speech dialogue models and full-duplex speech dialogue models have become pivotal new focuses for speech research. This section delineates the architectural paradigms, training methodologies, streaming capabilities, full-duplex strategies, and evaluation benchmarks that define the current state of this evolving domain.

\subsection{Architecture}

\subsubsection{Comprehension}

A fundamental requirement of speech interaction models is the ability to robustly comprehend spoken input, which hinges on effective speech encoding and seamless integration with the language model architecture. A common approach involves the utilization of dedicated speech encoders to transform raw audio into representations amenable to LLM. Whisper \cite{whisper} has become a popular choice for its strong speech modeling capabilities, and is integrated into several leading frameworks, including LLaMA-Omni \cite{fang2025llamaomni}, Mini-Omni \cite{xie2024miniomni}, Qwen2.5-Omni \cite{xu2025qwen2}, and Kimi-Audio \cite{kimiaudio}. Other self-supervised learning models, such as HuBERT \cite{hubert} and Wav2vec 2.0 \cite{w2v2}, also play a vital role in producing discrete, high-level speech units, which are foundational to ``textless NLP'' speech language models like GSLM \cite{gslm}, USDM \cite{kim2024paralinguisticsaware}, and Spirit-LM \cite{spiritlm}.

To bridge the modality gap between speech and text, these speech features are integrated into LLMs using modality adapters, also known as connectors. For systems utilizing discrete speech tokens, no complex connector is necessary, as these tokens are already aligned with the LLM's token-based input format. For models operating on continuous speech embeddings, most frameworks adopt a simple MLP adapter \cite{fang2025llamaomni,xie2024miniomni,fu2024vita} to project embeddings into the LLM's latent space. However, one key challenge lies in the length mismatch between speech and text sequences. The speech sequence length usually is much longer than that of the text sequence. For example, Whisper encoder produces audio embeddings at 50 frames per second, while a typical sentence contains fewer than 50 text tokens. This disparity necessitates downsampling to improve token density and computational efficiency. A common technique involves concatenating multiple consecutive embeddings. Some models, like Minmo \cite{chen2025minmo} and Freeze-Omni \cite{wang2025freezeomni}, implement learnable convolutional neural networks (CNNs) for more adaptive downsampling. IntrinsicVoice \cite{zhang2024intrinsicvoice} introduces a novel connector, GroupFormer, designed to intelligently compress speech sequences without losing critical information. To further enhance speech-text alignment, OmniDRCA \cite{tan2025omnidrca} proposes Contrastive Cross-modal Alignment, which optimizes the mutual distances between grouped speech and text representations, fostering semantic consistency prior to LLM processing.

Recently, increasing attention has been devoted to equipping speech interaction LLMs with reasoning or ``thinking'' abilities to enhance their overall intelligence. However, the dynamics of speech-based dialogue differ fundamentally from those of text-based conversations. For speech interaction LLMs, one major advantage lies in their potential to enable natural, fluid, and real-time communication, yet generating long-form internal reasoning often disrupts this natural flow, leading to perceptible delays. To make the conventional ``listen–think–speak'' mechanism more compatible with spoken interaction, several recent studies have proposed novel strategies that can be broadly categorized into two paradigms. The first, ``thinking while listening'' \cite{shih2025can,wu2025chronological,yu2025salmonn}, allows the model to produce short or truncated CoT thinking content during the user's speech input. This approach utilizes the otherwise idle listening period for internal reasoning, introducing little to no additional latency before the model's verbal response. The second paradigm, ``thinking while speaking'' \cite{chiang2025stitch,xie2025minireason,wu2025mind}, interleaves thinking tokens and speech tokens during streaming generation. By leveraging redundant inference time, it performs reasoning concurrently with speech output, maintaining real-time responsiveness. A persistent challenge for speech interaction LLMs is balancing answering accuracy and response latency. While extended reasoning sequences can improve task performance, they also introduce undesirable delays that disrupt conversational naturalness. Conversely, limiting reasoning to short or partial forms preserves responsiveness but constrains the model's thinking depth. Striking an optimal balance remains an open research problem.

\subsubsection{Generation}

For generation, models typically rely on either discrete speech tokens or continuous embeddings to guide the synthesis process. Discrete speech tokens offer a concise solution to enable LLMs with speech understanding and generation abilities by simply expanding the output vocabulary. In speech dialogue scenarios, where models must generate both text and speech tokens simultaneously, two common strategies emerge: using multiple output heads to generate separate token streams, exemplified by Moshi \cite{defossez2024moshi} and SALM-Duplex \cite{hu2025efficient}, or producing interleaved sequences of text and speech tokens, as seen in Spirit-LM \cite{spiritlm} and GLM-4-Voice \cite{glm4voice}. Another option is feeding LLM embeddings to the generation module \cite{yu2025salmonn,fang2025llamaomni,xu2025qwen2,chen2025minmo}. This module can either be trained from scratch \cite{fang2025llamaomni} or fine-tuned from a pre-trained TTS model, replacing the typical text input with LLM-derived embeddings \cite{yu2025salmonn}. In principle, the generation module can adopt any TTS architecture. However, a common choice is the autoregressive transformer, which generates semantic speech tokens followed by a flow-matching decoder to convert these tokens into spectrograms. This architecture is employed by models such as Seed-TTS \cite{anastassiou2024seed} and CosyVoice \cite{cosyvoice}. To enhance both speed and quality, recent advancements such as multi-token prediction (MTP) have been introduced in models like VocalNet \cite{wang2025vocalnet} and VITA-Audio \cite{long2025vita}. Additionally, \cite{arora2025chain} explores a Chain-of-Thought (CoT) paradigm in speech dialogue systems, where each conversational turn is structured into a pipeline of ASR transcription, textual response generation, and speech synthesis, thereby improving semantic coherence and response quality.

\subsection{Training}

The initial challenge in training a speech interaction LLM is the scarcity of large-scale spoken dialogue datasets. Existing resources such as IEMOCAP \cite{busso2008iemocap} and Fisher \cite{cieri2004fisher} often suffer from poor recording quality and require extensive preprocessing to filter out low-quality samples and eliminate background noise. To address this data scarcity, synthetic dialogue generation has become a widely adopted solution. Open-source synthetic datasets include VoiceAssistant-400K from Mini-Omni \cite{xie2024miniomni} and UltraChat from SLAM-Omni \cite{chen2024slam}. These datasets are typically constructed using powerful LLMs such as GPT-4 or LLaMA-3-70B-Instruct, which are employed to craft diverse conversational scenarios, including rewriting instructions into natural speech patterns, adding fillers, converting non-textual symbols into spoken equivalents and generating concise, speech-friendly responses \cite{fang2025llama,yu2025salmonn}. Advanced TTS models, including Bark TTS \footnote{\url{https://github.com/suno-ai/bark}} and CosyVoice 2 \cite{du2024cosyvoice2}, are then employed to convert these textual dialogues into extensive speech-to-speech QA pairs and multi-round conversations, thereby creating the necessary large-scale training corpora for these models \cite{veluri-etal-2024-beyond,chen2025emova}. Some models, like Moshi \cite{defossez2024moshi}, even develop TTS models using a single speaker's voice to ensure consistent acoustic identity throughout interactions.

The training of speech interaction LLMs typically involves sophisticated multi-stage strategies designed to bridge the modality gap between speech and text while preserving the LLM's intrinsic knowledge and reasoning abilities, as it is challenging to simultaneously enable speech understanding, speech generation, and dialogue management in a single pass. For codec-based models such as SpiritLM \cite{spiritlm}, SyncLLM \cite{veluri-etal-2024-beyond}, Moshi \cite{defossez2024moshi}, and GLM-4-Voice \cite{glm4voice}, training generally follows two major phases. The first phase involves pre-training on large-scale speech or speech-text corpora to teach the model to generate speech tokens reliably. For instance, Moshi leverages over 7 million hours of audio, while GLM-4-Voice utilizes approximately 500 billion speech tokens. Interleaving text and speech during pre-training is shown to enhance modality alignment \cite{spiritlm,glm4voice}, and incorporating text-only data helps mitigate catastrophic forgetting \cite{glm4voice}. The second phase focuses on fine-tuning the pre-trained model on dialogue-specific data, allowing the model to develop fluent and contextually appropriate voice interaction capabilities.

For models employing a downstream generation module such as Freeze-Omni \cite{wang2025freezeomni}, Minmo \cite{chen2025minmo}, Qwen2.5-Omni \cite{xu2025qwen2} and SALMONN-omni \cite{yu2025salmonn}, the training strategy is also typically two-phased. The first stage connects a speech encoder for comprehension tasks, while the second integrates a generation module for producing output speech. In most cases, the initial training phase excludes the generation component, focusing on optimizing the encoder and connector. During the second stage, the speech encoder and connector are often frozen to improve training efficiency when connecting the generation module. Notably, SLAM-Omni \cite{chen2024slam} distinguishes itself by achieving competitive performance with a single-stage training approach, directly training on speech-to-speech interaction tasks. Recently, RL methods, particularly DPO, has been incorporated into training pipelines. These methods aim to improve full-duplex dialogue modeling \cite{yu2025salmonn} and better align model behavior with user preferences \cite{wu2025aligning,lin2024align}.

\subsection{Streaming}

\subsubsection{Streaming Input}

The requirement for models to spontaneously process speech input and generate responses, particularly in full-duplex conversational scenarios, mandates specialized designs for handling streaming input. Various strategies have been developed to integrate these streaming speech inputs into LLMs. One approach involves cross-attention mechanisms where the LLM processes incoming speech embeddings in a chunked or step-wise fashion, as exemplified by the wait-k policy, which defines a fixed pre-decision ratio for processing speech embedding steps before predicting subword units \cite{chen2024bestow}. Another prominent strategy entails concatenating different inputs, where models integrate speech by prepending speech prompts to text prompts, or by directly feeding continuous speech embeddings from the speech encoder into the LLM. Moshi \cite{defossez2024moshi} and Mini-Omni \cite{xie2024miniomni} are notable examples of models that adopt this direct integration, enabling the LLM to understand speech instructions without a prerequisite text transcription. Furthermore, some models implement interleaving of different inputs, where LLM text response tokens are interleaved with environmental and assistant stream embeddings into a single sequence, allowing the LLM backbone to model them jointly in an autoregressive manner. This joint processing, as seen in models like SyncLLM \cite{veluri-etal-2024-beyond} and SALMONN-Omni \cite{yu2025salmonn}, is particularly effective for full-duplex interaction as it inherently accommodates complex conversational dynamics, including overlapping speech and interruptions, by treating user and system audio streams simultaneously.

\subsubsection{Streaming Output}

The co-generation of speech and text is a critical aspect of streaming output, and models employ diverse methods to achieve this synchronicity. Some architectures, such as SpiRit-LM \cite{spiritlm} and GLM-4-Voice \cite{glm4voice}, are trained on interleaved speech and text data, enabling them to generate content in either modality. Conversely, models like PSLM \cite{mitsui-etal-2024-pslm} and Mini-Omni \cite{xie2024miniomni} adopt a parallel generation paradigm, directly decoding both text and speech tokens simultaneously to significantly reduce latency. This parallel processing is further extended by Moshi \cite{defossez2024moshi}, which uses a multi-stream architecture to jointly model input and output audio streams. Additionally, approaches like OmniDRCA \cite{tan2025omnidrca} fuse speech and text representations for joint autoregressive modeling, ensuring temporal alignment during generation. A prevalent strategy involves leveraging the LLM's output hidden states to guide speech synthesis, often through a dedicated streaming speech decoder or synthesizer. This allows models such as LLaMA-Omni \cite{fang2025llamaomni}, Freeze-Omni \cite{wang2025freezeomni}, and Qwen2.5-Omni \cite{xu2025qwen2} to extend the LLM's textual intelligence to the speech modality while maintaining low latency.

The inherent frequency mismatch between text and speech necessitates robust alignment strategies during co-generation. Many models employ fixed alignment mechanisms, such as periodic synchronization or pre-defined chunk sizes, to ensure a smooth interplay between audio and text streams. For instance, OmniFlatten \cite{zhang2024omniflatten} and Qwen2.5-Omni \cite{xu2025qwen2} define specific chunk sizes for text and speech tokens and interleave them into a single flattened sequence for training and real-time streaming output. SALMONN-Omni \cite{yu2025salmonn} similarly employs a periodic synchronization mechanism, processing fixed durations of input speech and generating matching durations of speech responses in time blocks. In contrast, certain approaches utilize dynamic alignment techniques. Moshi \cite{defossez2024moshi} integrates temporal alignment between speech and its transcript to enable modality switching and consistent internal representations.

\subsection{Full-Duplex Strategies}

Full-duplex conversation, characterized by simultaneous bidirectional communication, is a critical feature for mimicking human-like interaction. A model is identified as ``full-duplex'' when the model can hear and speak at the same time, which means the model can go beyond turn-based conversations, modeling more complex interactions like barge-in (user's interruption) and backchanneling (e.g., acknowledgments like ``uh-huh''). Various strategies have emerged to achieve this in cascaded dialogue systems \cite{wang2024full,liao2025flexduo,zhang2025llm}. The cascaded approach, while traditional, still forms a baseline, relying on separate modules including VAD (Voice Activity Detection), ASR, LLM and TTS. However, it is worth noting that most speech interaction models nowadays are still turn-based, meaning they cannot listen while generating speech. For these turn-based models, incorporating the VAD module can also achieve full-duplex dialogue interaction \cite{kimiaudio}. Despite their utility in simpler exchanges, modular architectures struggle with the complexity of real-world, fluid conversation, particularly with content-sensitive barge-ins and nuanced backchannel cues.

For codec-based full-duplex speech interaction models, full-duplex capability is achieved through perpetual speaking. 
These models continuously receive incoming speech while autoregressively generating speech tokens, producing silence where appropriate, without explicitly modeling turn-taking. 
In essence, they are ``always listening and always speaking''. 
Moshi \cite{defossez2024moshi} is a pioneering example. Its multi-stream audio language model handles both input and output streams as unified autoregressive token sequences, eliminating discrete speaker turns and naturally accommodating overlap and interruptions. Similarly, OmniFlatten \cite{zhang2024omniflatten} supports continuous, bidirectional interaction through a chunk-based flattened stream that merges speech and text tokens into a single sequence. This intertwined generation mechanism enables a more fluid and realistic conversational flow, addressing the artificial constraints of turn-based systems.

For models that do not operate on a codec-based token stream, explicit turn-taking modeling is necessary to achieve full-duplex functionality. One strategy, known as the model-as-server approach, runs two interdependent LLM processes, one for listening and one for speaking. This technique is employed by models like VITA \cite{fu2024vita} and Freeze-Omni \cite{wang2025freezeomni}, enabling simultaneous comprehension and generation. However, this dual-LLM setup introduces substantial computational and memory overhead, as both instances must operate in parallel. Another elegant solution is offered by the standalone state prediction strategy, exemplified by SALMONN-Omni \cite{yu2025salmonn}. Unlike codec-based systems, SALMONN-Omni does not inject audio tokens directly into the LLM's input space. Instead, it introduces a thinking mechanism within a single LLM process, enabling the model to manage transitions between listening and speaking autonomously. This is accomplished by training the LLM to generate state transition tokens as part of its output sequence, eliminating the need for separate full-duplex predictors or multiple LLMs.

\subsection{Evaluation}

The comprehensive evaluation of speech interaction LLMs requires a nuanced and multifaceted approach that accounts for both linguistic accuracy and the subtleties of spoken communication. A growing number of benchmarks have been proposed to assess these models across diverse interactional dimensions \cite{chen2024voicebench,li2025baichuan,gao2024benchmarking,cui2025voxeval,ahelm}. Foundational efforts such as VoiceBench \cite{chen2024voicebench} and OpenAudioBench \cite{li2025baichuan} primarily evaluate conversational capabilities in single-turn interactions, spanning tasks related to general knowledge, instruction following, and safety alignment. These evaluations are largely semantic-focused: generated speech is transcribed into text, and the assessments are then performed in the text domain. Recent benchmarks have expanded the scope of evaluation toward more comprehensive and realistic scenarios. URO-Bench \cite{yan2025uro} introduces multi-turn dialogue evaluation, while VocalBench \cite{liu2025vocalbench} and SOVA-Bench \cite{hou2025sova} incorporate paralinguistic assessments, evaluating aspects such as emotion, prosody, and speaker traits. Talking Turns \cite{arora2025talking} and Full-Duplex-Bench \cite{lin2025full} and specifically assess full-duplex conversational behaviors, including pause handling, turn-taking, barge-ins, and backchanneling, using a mix of automatic metrics to quantify real-time interaction dynamics. Additionally, S2S-Arena \cite{jiang2025s2s} proposes an arena-style benchmark for speech-to-speech evaluation, where human judges perform pairwise comparisons of dialogue quality, offering a richer and more holistic evaluation.

In terms of evaluation metrics, speech interaction LLMs are assessed using a mix of objective and subjective measures. Standard metrics include WER and CER for speech recognition and synthesis accuracy, UTMOS \cite{saeki2022utmos} for predicted speech quality, and F1-scores for evaluating response styles or intent matching. For conversational behavior, Takeover Rate (TOR) is employed to measure the accuracy of turn taking and barge-in, and latency is calculated to evaluate the real-time capability to process different user requests.
Importantly, a growing number of benchmarks now employ advanced LLMs such as GPT-4o as automated judges to evaluate criteria like helpfulness, fluency, coherence, relevance, engagement, factual correctness, and instruction adherence. This shift toward LLM-as-a-judge evaluation reflects a broader trend in the field, moving toward scalable and comprehensive assessment methodologies for next-generation speech interaction systems.

\section{Audio-Visual modeling with LLMs}

\begin{figure}[ht]
    \centering
    \includegraphics[width=\linewidth]{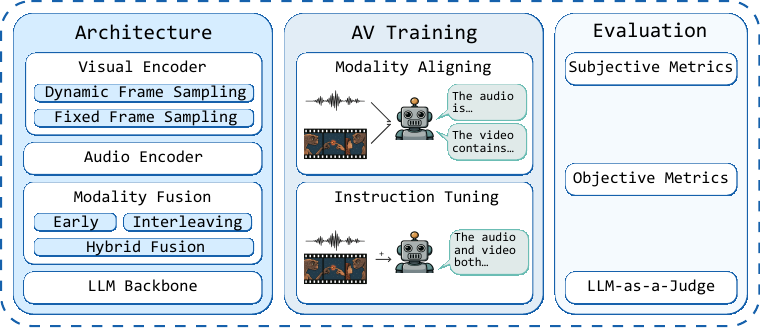}
    \caption{An overview of audio-visual modeling with LLMs, highlighting modular architectures with multimodal fusion, multi-stage training, reinforcement learning optimization, and evaluation grounded in standard task metrics.}
    \label{fig:chpt6}
\end{figure}

\begin{figure*}[h]
    \centering
    \includegraphics[width=\linewidth]{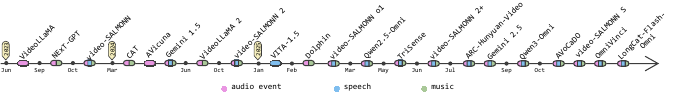}
    \caption{Timeline of recent Audio-Visual LLMs. Some works not mentioned in the main text are cited here \cite{team2024gemini, comanici2025gemini, chen2025avocado, sun2025videos, ye2025omnivinci, longcat-flash-omni}.}
    \label{fig:audio_visual_timeline}
\end{figure*}

The dynamic interplay between visual and auditory information is fundamental to human perception, offering complementary insights that enrich our understanding of the world. Similarly, in the realm of Multimodal LLMs (MLLMs), integrating audio with visual data has emerged as a crucial area of research in video understanding, moving beyond single-modality approaches to achieve a more comprehensive and accurate interpretation of complex, real-world scenarios. This section delves into the key components, training methodologies, and evaluation strategies employed in advancing audio-visual modality integration within MLLMs.

\subsection{Architecture}

The effective integration of audio and visual modalities in MLLMs hinges on several critical components, including complex feature extraction methods, diverse fusion strategies, and tailored model architectures, often complemented by dedicated preprocessing steps. 
% \subsubsection{Feature extraction} typically involves specialized encoders for each modality. For visual inputs, pre-trained vision transformers (ViTs) and CLIP-based models, such as CLIP ViT-L/14 or ViT-G, are widely utilized to extract visual embeddings or frame-level features \cite{tang2025empowering, chen2023vast, fu2024vita, damonlpsg2024videollama2, 10.5555/3692070.3693991}. 
% These encoders process images or sequences of video frames, with approaches like sampling a fixed number of frames or dynamically adjusting frame rates to capture temporal dynamics \cite{fu2024vita, zhang-etal-2023-video, damonlpsg2024videollama2}. 
% For audio, prominent encoders include BEATs \cite{10.5555/3692070.3693991, damonlpsg2024videollama2}, Whisper \cite{10.5555/3692070.3693991, damonlpsg2024videollama2, guo2025aligned}, CLAP \cite{tang2025empowering}, and the audio encoder of ImageBind \cite{han2023imagebind, zhang-etal-2023-video, su-etal-2023-pandagpt}. 
% These convert raw audio waveforms or mel-spectrograms into dense vectors, capturing auditory features and temporal dynamics. 

\subsubsection{Feature extraction} This typically involves specialized encoders for each modality. For visual inputs, pre-trained vision transformers, such as CLIP \cite{radford2021learning} or SigLIP \cite{zhai2023sigmoid}, are widely used to extract visual embeddings or frame-level features. 
These encoders process images or sequences of video frames, with approaches such as sampling a fixed number of frames or sampling frames at a certain frame rate. 
For audio, pre-trained encoders including BEATs \cite{beats} for audio event processing and Whisper \cite{whisper} for speech signal processing are commonly used \cite{sun_video-salmonn-o1_2025,xu2025qwen2}. 
These audio encoders convert raw audio waveforms or Mel-spectrograms into dense vectors, capturing auditory features and temporal dynamics. 

\subsubsection{Preprocessing steps} These are essential to standardize inputs; for instance, audio signals are often resampled and transformed into mel-spectrograms, while video frames undergo resizing and normalization.

\subsubsection{Fusion strategies} These are used to determine how audio and visual features are combined, ranging from naive concatenation to interleaved approaches. 
Naive concatenation refers to concatenating tokens of each modality to form a multimodal token sequence that serves as the LLM input. For instance, LLaMA-AVSR \cite{cappellazzo2025large}, Video-LLaMA \cite{zhang-etal-2023-video} and VideoLLaMA 2 \cite{damonlpsg2024videollama2} employ this naive concatenation approach.
Other models like video-SALMONN \cite{10.5555/3692070.3693991} and ARC-Hunyuan-Video \cite{ge2025arc} align audio-visual tokens by concatenating or adding them in the temporal dimension, thereby achieving precise temporal alignment of audio and video information.
Interleaved fusion, a more sophisticated approach, involves orchestrating the temporal relationship of tokens from audio and video by creating interleaved sequences. This also ensures temporal synchronism and fine-grained alignment between the modalities. 
Models using interleaved fusion include AVicuna \cite{tang2024avicuna}, video-SALMONN 2 \cite{tang2025video}, Qwen2.5-Omni \cite{xu2025qwen2}, et al.

\subsubsection{Model architectures} It typically consists of multimodal encoders, projection layers (or adapters), and an LLM backbone. The multimodal encoders and the LLM are always well-pre-trained, so the main architectural differences between different models are mainly reflected in the projection modules. The projection modules, often referred to as ``connectors" or ``aligners", bridge the gap between modality-specific feature spaces and the LLM's token embedding space. The multilayer perceptron (MLP) is a common structure to serve as the connector, especially for models that separately process audio features and visual features \cite{zhang-etal-2023-video, xu2025qwen2, tang2025video}. 
The Q-Former is also an option, which learns reasonable query embeddings that are understandable by the LLM. For instance, video-SALMONN \cite{10.5555/3692070.3693991} designs a multi-resolution causal Q-Former to connect pre-trained audio-visual encoders and the backbone large language model. Some models introduce specialized modules like the ``audio-visual multi-scale adapter'' of Dolphin \cite{guo2025aligned} for comprehensive and accurate understanding across temporal and spatial dimensions. Others, like CAT \cite{ye2024cat}, design a ``clue aggregator'' to dynamically capture question-aware visual and audio hidden features, enriching the detailed knowledge for the LLM. 
% The Thinker-Talker architecture, used in Qwen2.5-Omni \cite{xu2025qwen2}, exemplifies a design where a ``Thinker'' LLM generates text, and a ``Talker'' model produces audio tokens, demonstrating a unified system for both perception and generation. 
% These components interact to transform raw multimodal inputs into a unified representation that the LLM can process, enabling it to generate coherent and contextually rich responses, enhancing overall multimodal performance by leveraging the complementary nature of audio and visual signals.

\subsection{Audio-Visual Training}

Training robust audio-visual MLLMs necessitates diverse approaches, often involving multi-stage pipelines and innovative techniques to overcome data limitations and modality inconsistencies. 
Many models adopt a multi-stage training process. 
Typically, the initial stages focus on modality-text alignment, where individual modality encoders (visual and audio) are aligned with the LLM using large-scale unimodal or dual-modal datasets. 
For example, VITA-1.5 \cite{fu_vita-15_2025} dedicates its first stage to vision-language training and the second to audio input processing. 
Video-LLaMA \cite{zhang-etal-2023-video} similarly pre-trains vision-related and audio-related components on large-scale caption datasets, initially even training the audio-language branch using visual-text data due to audio-text data scarcity.

Subsequent stages typically involve multimodal instruction tuning or joint training on carefully curated datasets to enhance combined audio-visual understanding and instruction-following capabilities. 
For instance, Dolphin \cite{guo2025aligned} uses an ``audio-visual understanding caption \& instruction-tuning dataset'' (AVU). 
Audio-Visual LLM employs a ``modality-augmented training'' (MAT) strategy, which involves integrating modality-specific tokens to selectively activate visual and/or auditory encoders, allowing for end-to-end joint training with visual-only, audio-only, and audio-visual data. 
This addresses the challenge of flexibly fusing different modalities within a single batch.

A crucial aspect of audio-visual training is the curation and generation of high-quality, large-scale multimodal datasets. 
Faced with the scarcity of audio-visual video datasets with precise temporal annotations, researchers have developed innovative methods to synthesize such data. 
Examples include PU-VALOR, derived from VALOR-32K by applying random temporal scaling and permutation to clustered videos. 
AVU \cite{guo2025aligned}, another significant dataset, comprises 5.2 million diverse, open-ended data tuples and employs a novel data partitioning strategy, including negative samples to mitigate hallucinations. 
Similarly, NExT-GPT \cite{wu2024next} introduces a ``modality-switching instruction tuning'' (MosIT) dataset, manually curated for complex cross-modal understanding and generation. 
CAT \cite{ye2024cat} collects AVinstruct, an ``audio-visual joint instruction dataset'', to enhance its capacity for cross-semantic correlations and address Audio-Visual Question Answering (AVQA) tasks. 
VAST-27M \cite{chen2023vast} is an automatically generated large-scale omni-modality video caption dataset, where LLMs integrate single-modality captions and subtitles into unified omni-modality captions. 
TriSense \cite{li2025watch} introduces TriSense-2M, a 2-million-sample dataset with event-based annotations across vision, audio, and speech, designed to support flexible modality combinations and long-form videos.

% Innovations in training techniques also include addressing specific challenges. 
Other innovations in training techniques are proposed to tackle specific challenges.
To counteract modality dominance and ensure balanced feature extraction, video-SALMONN \cite{10.5555/3692070.3693991} proposes a diversity loss and an unpaired audio-visual mixed training scheme. 
This enforces the model to extract information from both audio and video inputs without over-relying on a single dominant modality, leading to improved audio-visual understanding and co-reasoning abilities. video-SALMONN 2 \cite{tang2025video} designs a new captioning metric and applies Multi-round Direct Preference Optimization (MrDPO)  to enhance captioning quality by optimizing for completeness and accuracy. This helps reduce hallucination and repetition rates in generated descriptions. video-SALMONN-o1 \cite{sun_video-salmonn-o1_2025} proposes process DPO to enhance the audio-visual reasoning capability of the model.
CAT \cite{ye2024cat} further proposes an ``AI-assisted Ambiguity-aware Direct Preference Optimization'' (ADPO) strategy to retrain models to favour non-ambiguous responses and improve localization. 

\subsection{Audio-Visual Evaluation}

Evaluating the effectiveness of audio-visual MLLMs involves multifaceted approaches, combining both qualitative and quantitative metrics to assess performance across various tasks, fusion effectiveness, robustness, and generalizability. 
Quantitative metrics are widely used to measure performance across specific tasks. 
For question answering (QA) tasks, accuracy is a primary metric. Traditional benchmarks mainly focus on content understanding, like MSRVTT-QA \cite{xu2017video}, AVSD \cite{alamri2019audiovisual}, MUSIC-AVQA \cite{li2022learning}, and ActivityNet-QA \cite{yu2019activityqa}. 
In recent years, many new audio-visual benchmarks have emerged, which focus not only on the content but also on deeper-level reasoning skills.
For instance, Video-MME \cite{fu2025video} is a well-annotated benchmark and includes videos of various domains and questions in a broad range.
AVUT \cite{yang2025audiocentric} targets audio-centric video understanding while addressing the ``text shortcut'' problem in multimodal evaluation. Daily-Omni \cite{zhou2025daily} evaluates the model's audio-visual reasoning performance across different temporal contexts. Video-Holmes \cite{cheng2025video} evaluates the model in complex video reasoning tasks.
For captioning tasks, traditional metrics like CIDEr \cite{vedantam2015cider} and SPICE \cite{anderson2016spice} scores are commonly reported to evaluate the quality and semantic relevance of generated captions, especially of short captions. For detailed caption evaluation, \cite{tang2025video} proposes a metric based on atomic events.
Audio-Visual Speech Recognition (AVSR) tasks typically use WER to assess transcription accuracy.

Despite significant advancements, current evaluation methodologies still face limitations and open issues. 
There remains a shortage of large-scale, high-quality, and fine-grained audio-visual datasets with detailed annotations, which hinders the comprehensive evaluation of fine-grained understanding. Accurate audio-visual synchronization also has very little data for training and evaluation.
In addition, existing datasets are often limited in scale and scope, limiting the development of more advanced multimodal reasoning capabilities. 
There is also a challenge in ensuring modality balance during training and evaluation, as models can sometimes default to unimodal shortcuts if one modality dominates, leading to unimodal biases. 
The need for more complex, reasoning-intensive tasks that demand deep contextual understanding and multi-step inference across modalities is also highlighted. 
The computational cost associated with processing long audio-visual sequences remains a practical challenge in training and inference, which affects how efficiently models can be evaluated.

\section{Challenges and Future Work}

\subsection{Audio Representation} 

A significant challenge lies in effectively converting diverse audio signals, including human speech, natural sounds, and music, into representations that LLMs can process. While discrete audio tokens derived from self-supervised speech encoders like HuBERT \cite{hubert} and wav2vec 2.0 \cite{w2v2} have shown promise, enabling LLMs to learn from raw audio without text or expert labels, issues persist with sequence length inconsistency and the optimal choice of encoding units. Future work needs to focus on robust methods for capturing both linguistic content and expressive paralinguistic features (like pitch and style) in these representations, ensuring they are efficient and avoid performance degradation seen with continuous speech features, while exploring prompt-aware mixture of audio encoders for task-specific feature emphasis. Additionally, there is a need to refine training to bridge the gap between speech and text performance, as current models may not perform as well in speech as their text-only counterparts.

\subsection{Audio Comprehension}

Current audio LLMs face significant hurdles in complex audio comprehension, primarily stemming from the relative scarcity of audio data compared to text, which limits the investigation of data scaling effects. This data limitation manifests in critical performance gaps, including deficient deductive reasoning, as observed in Audio Entailment tasks \cite{deshmukh2025audio}, and a propensity for object hallucination when asked to discriminate specific sounds within a scene \cite{kuan2024understanding}. Furthermore, their deep reasoning capabilities remain underdeveloped, often failing to follow complex logical chains that are intuitive to humans. Future work must therefore prioritize not only scaling and diversifying audio datasets but also enhancing model architecture and training methodologies. Key research directions include developing models that integrate multimodal context, such as geographical and cultural knowledge, exploring reinforcement learning to improve question-answering performance, and pushing the frontier of model capabilities to include robust generalization, effective in-context learning, and sophisticated multi-step reasoning for novel and more demanding audio understanding tasks.

\subsection{Audio Generation} 

The frontier of audio generation with LLMs faces several intricate challenges that necessitate continued research. A significant hurdle lies in consistently achieving high-fidelity and universally natural-sounding audio generation, particularly for diverse and complex soundscapes beyond human speech, such as music, singing, and varied environmental sounds. Future work aims to move beyond pipelines that rely on intermediate text-based transcriptions to enable more seamless and expressive direct audio outputs. Another critical area is gaining fine-grained controllability and expressivity over generated audio, including the ability to precisely control emotions, intonations, speaking styles, timbres, and accents, while also preserving speaker identity across different generated utterances. Generating coherent long-form audio that maintains semantic and acoustic consistency over extended durations, such as multi-minute spoken narratives, remains a complex task due to the high temporal resolution of audio tokens and associated memory constraints. Researchers are also focused on developing unified models capable of generating diverse audio types (speech, music, sounds) and seamlessly integrating with other modalities like image and video generation within a single, cohesive framework. Finally, addressing the ethical considerations of generative audio, such as mitigating the potential for malicious content creation or voice impersonation, will require ongoing development of robust safety mechanisms and watermarking techniques.

\subsection{Speech Interaction} 

The domain of speech interaction with LLMs is rapidly evolving, yet it presents distinct challenges and avenues for future exploration. A primary research direction focuses on achieving truly natural and low-latency full-duplex spoken dialogue, which necessitates breakthroughs in managing complex conversational dynamics such as effective turn-taking, backchanneling, handling overlapping speech, and context-dependent barge-in, moving beyond conventional half-duplex systems. The robustness of models to real-world audio conditions, including noise, varying speaker characteristics, and linguistic disfluencies, remains suboptimal. The scarcity and qualitative limitations of suitable training data pose another pervasive issue; specifically, there is a dearth of large-scale datasets that capture varied speaking styles and diverse real-world conversational scenarios. Finally, the field necessitates the establishment of unified, reproducible, and comprehensive evaluation benchmarks that extend beyond text-based metrics to rigorously assess full-duplex capabilities, paralinguistic understanding, and generation quality in diverse real-world contexts.

\subsection{Audio-Visual Comprehension} 

The integration of audio and visual modalities for a holistic understanding of dynamic scenes and complex events in videos presents distinct challenges. Key difficulties include achieving precise temporal alignment and fusion of information between audio and visual streams, mitigating modality bias where models might over-rely on one modality, and handling noisy labels in weakly-supervised settings. Existing Multimodal LLMs often struggle to discern subtle relationships and exhibit hallucinations due to their limited capacity to perceive complex multimodal signals and their interrelationships. Future research needs to focus on novel architectures like multi-resolution causal Q-Formers and multi-scale adapters to improve fine-grained spatial and temporal alignment. Developing high-quality audio-visual instruction datasets and applying reinforcement learning frameworks are crucial to enhance cross-modal reasoning and mitigate hallucinations, allowing models to understand audio-centric video information comprehensively.

The integration of audio into large language models marks a decisive step toward more human-like and embodied artificial intelligence. Looking ahead, progress will depend not only on technical advances but also on the way these systems are applied and governed. We highlight two intertwined directions that shape the field's trajectory.

\subsection{Applications with societal value}

Audio-native intelligence is poised to transform several domains. In healthcare, speech and non-verbal cues provide biomarkers for conditions such as depression, autism, and cognitive decline, enabling scalable tools for early screening and digital phenotyping. In education, real-time spoken dialogue systems can democratize access to personalized tutoring and language learning. In culture and creativity, expressive audio generation fosters human–AI co-creation in music, entertainment, and digital companionship. In robotics, auditory perception enhances environmental awareness and enables natural interaction, supporting embodied AI in daily life.

\subsection{Ethical, safety, and governance considerations}

These opportunities are counterbalanced by risks. Voices are biometric identifiers; unauthorized cloning and inference threaten privacy and security. Current systems remain biased toward high-resource languages and standardized accents, risking exclusion of underrepresented communities. The growing realism of synthetic voices intensifies threats of fraud and disinformation. Addressing these challenges requires safeguards such as watermarking and misuse detection, transparent documentation of datasets, and cross-sector standards for responsible deployment.

\bibliographystyle{IEEEtran}
\bibliography{bib/chpt3,bib/chpt4,bib/chpt5,bib/chpt6}

% \newpage

%\section{Biography Section}
%If you have an EPS/PDF photo (graphicx package needed), extra braces are needed around the contents of the optional argument to biography to prevent the LaTeX parser from getting confused when it sees the complicated $\backslash${\tt{includegraphics}} command within an optional argument. (You can create your own custom macro containing the $\backslash${\tt{includegraphics}} command to make things simpler here.)
 
% \vspace{11pt}

% \bf{If you include a photo:}\vspace{-33pt}
% \begin{IEEEbiography}[{\includegraphics[width=1in,height=1.25in,clip,keepaspectratio]{fig1}}]{Michael Shell} 
% Use $\backslash${\tt{begin\{IEEEbiography\}}} and then for the 1st argument use $\backslash${\tt{includegraphics}} to declare and link the author photo. Use the author name as the 3rd argument followed by the biography text.
% \end{IEEEbiography}

% \vspace{11pt}

% \bf{If you will not include a photo:}\vspace{-33pt}
%\begin{IEEEbiographynophoto}{John Doe}
%Use $\backslash${\tt{begin\{IEEEbiographynophoto\}}} and the author name as the argument followed by the biography text.
%\end{IEEEbiographynophoto}

%\vfill

\end{document}